
\magnification 1200
\baselineskip 24 pt
\def \la {\langle}
\def \ra {\rangle}

\line {\hfill March 1994}
\vskip 3 true cm
\centerline {\bf BELL'S INEQUALITIES AND DENSITY MATRICES.}
\centerline{\bf REVEALING
``HIDDEN" NONLOCALITY.}

\vskip 2 true cm

\centerline {Sandu Popescu\footnote{*}{present address: Tel Aviv
University, School of Physics and Astronomy, Ramat Aviv, Tel Aviv,
Israel.}}
\vskip 2 true cm
\centerline{Department of Physics,}
\centerline{ Technion - Israel Institute of Technology}
\centerline{32000 Haifa, Israel }
\vskip 3 true cm
\centerline{Abstract}

As is well known, quantum mechanical behavior cannot, in general, be
simulated by a local hidden variables model. Most -if not all- the proofs
of this incompatibility refer to the correlations which arise when each of
two (or more) systems separated in space is subjected to a single ideal
measurement.  This setting is good enough to show contradictions between
local hidden variables models and quantum mechanics in the case of pure
states.  However, as shown here, it is not powerful enough in the case of
mixtures. This is illustrated by an example. In this example, the
correlations which arise when each of two systems separated in space is
subjected to a single ideal measurement are classical; only when each
system is subjected to a {\it sequence} of ideal measurements non-classical
correlations are obtained. We also ask
 whether there are situations for which even this last procedure is not
powerful enough and non-ideal measurements have to be considered as well.
\vfill
\eject

As it is well known, when measurements are performed on two quantum systems
separated in space their results are correlated in a manner which, in
general, cannot be explained by a local hidden variables model. But 30
years after Bell's pioneering paper [1] we still lack a complete
classification of quantum states into local and nonlocal ones. While the
case of pure states is completely solved [2], for density matrices only
partial results have been obtained so far [3]. In this letter I show that
for solving this problem we have to change the usual way we think at Bell's
inequalities and local hidden variable models.

Bell's original proof and most -if not all- the subsequent alternative
proofs [4] (with or without inequalities) have a common aspect: They
consider the case in which each of the two systems is subjected to a {\it
single ideal} local measurement (chosen at random among many possible ideal
measurements). For example in the case of spin 1/2 particles one usually
considers that each particle is subjected to a single Stern-Gerlach
measurement, that is, to a measurement of the spin along some arbitrary
direction. More general, by ideal measurements I mean measurements as
defined in the postulates of quantum mechanics [von Neumann, Dirac], that
is, when a system is described by an n-dimensional Hilbert space, the
observables to which measurements refer correspond to hermitian operators
acting on this n-dimensional space, and the only possible results of a
measurement are the eigenvalues of the measured operator. After such a
measurement the state of the measured system becomes the projection of the
initial state onto the subspace corresponding to the observed
eigenvalue. Of course in principle one could consider that each of the two
space separated systems is subjected to much more complicated experiments,
that is, to a {\it sequence} of {\it non ideal} measurements instead of a
single ideal one. Clearly, to prove that a given quantum state is nonlocal,
it is enough to show that it violates some usual Bell inequality (obtained
by considering a single ideal measurement performed on each side). On the
other hand, to prove that a quantum state is local, one must show that the
correlations between the results of {\it any} local experiments can be
described by a local hidden variables model.

The above being said, the question still remains whether considering
general local measurements gives us essential new information about the
locality or nonlocality of quantum states.  In this letter I will show that
the answer is to this question is ``yes" .

To put things in a better perspective, let us recall first the status of
pure states. Although it might well be that considering general
measurements will yield new and better inequalities, the basic question of
which pure states are classical or not can be answered without going beyond
the usual scheme. Indeed, as is well known [2] every entangled pure state
violates some usual Bell inequality and it is therefore nonlocal. The only
pure states which do not yield nonlocal correlations when a single ideal
measurement is performed on each particle are the direct product states and
they are obviously local. On the other hand, as I will show, the situation
turns out to be different when one deals with mixed states. More precisely,
I will give an example in which if each of two systems separated in space
is subjected to a single ideal measurement the resulting correlations are
classical but if each system is subjected to more complicated experiments
the resulting correlations are nonlocal. The ``more complicated
experiments" I consider here are sequences of two ideal measurements.

The density matrices I analyze in this paper were discovered by Werner
[5]. Consider two systems separated in space, each system living in a $d$
dimensional Hilbert space, and let the (mixed) state of the two systems be

$$ W={1\over {d^2}}\bigl({1\over d}I^{(d\times
d)}+2\sum_{i<j;i,j=1}^{d}\vert S_{ij}\ra\la S_{ij}\vert\bigr),\eqno(1) $$
where $I^{(d\times d)}$ is the identity matrix in the $d\times d$
dimensional space of the two systems and $\vert S_{ij}\ra$ is the ``spin
1/2 singlet" state
$$ \vert S_{ij}\ra={1\over {\sqrt 2}}\bigl(\vert i\ra_1\vert j\ra_2-\vert
j\ra_1 \vert i\ra_2\bigr),\eqno(2) $$
where $\{\vert i\ra_1\}$ and $\{\vert j\ra_2\}$ denote orthogonal bases in
the hilbert spaces of particle 1 and 2 respectively. (The form (1) of the
matrix W can be easily obtained from the original form given by Werner by
noting that the ``flip" operator V which appears in [5] can be expressed as
$V=I-2\sum_{i<j}\vert S_{ij}\ra\la S_{ij}\vert$.)

An important property of these density matrices is that they cannot be
decomposed into mixtures of pure direct product states. As direct products
are the onlyclassical pure states, and as Werner's matrices are not
mixtures of such states, one feels intuitively that these density matrices
are nonclassical. However Werner has proven that these density matrices do
not violate any standard Bell inequality. That is, when each system is
subjected to a single ideal measurement the resulting correlations are
classical. (Werner has explicitly writtena local hidden variables model
which simulates these correlations.)  But is there a local hidden variables
model which can simulate the results of {\it any} measurements performed on
the two systems prepared in such a mixed state?

Very recently progress has been made in understanding the properties of
these strange states. It has been shown [6] that when two spin 1/2
particles are in such a state (Werner's $2\times 2$ dimensional matrix)
they form a ``quantum channel" which can be used for (imperfect)
teleportation. Thus the $2\times 2$ dimensional Werner matrix has indeed
nonclassical aspects. However teleportation does not involve local
measurements performed on each of the spins outside the light cone of each
other, so from the fact that such a matrix can be used for teleportation we
do not learn immediately anything about the nonlocal correlations it might
generate. Nevertheless this suggests that we should consider more carefully
the question of nonlocality of Werner type matrices and go beyond the usual
scheme in which each system is subjected to a single ideal measurement.

In this letter I will consider Werner type matrices of dimension $\geq
5\times 5$. Suppose now that each of the two particles is subjected to two
consecutive ideal measurements. First each particle is subjected to a
measurement of a 2-dimensional projection operator,
$$
P={\vert 1\ra_1}{_1\la 1\vert}+{\vert 2\ra_1}{_1\la 2\vert} \eqno(3)
$$
on particle 1 and
$$
Q ={\vert 1\ra_2}{_2\la 1\vert}+{\vert 2\ra_2}{_2\la 2\vert} \eqno(4)
$$
on particle 2.
{\it After} the first measurement is performed and its result is registered
an observer situated near particle 1 chooses at random to measure one of
the two operators A or A', whose exact form will be described below. It is
important to note that the decision which operator to measure is taken only
after the measurement of P is completed. Similarly, an observer situated
near particle 2 chooses at random between a measurement of B or B'. This
scheme is almost identical to the one usually used for deriving Bell's
inequalities with the difference that A, A', B and B' are not measured
directly on the particles in the original state but after the measurements
of P and Q respectively.

The operators A, A', B and B' have each three different eigenvalues, 1, -1
and 0. The eigenvalues 1 and -1 are nondegenerate and the corresponding
eigenstetes belong to the subspaces $\{ \vert 1\ra_1, \vert 2\ra_1\}$ and
$\{ \vert 1\ra_2, \vert 2\ra_2\}$ respectively. The eigenvalue 0 is highly
degenerate and corresponds to the rest of the hilbert spaces, that is to
$\{ \vert 3\ra_1,..., \vert d\ra_1\}$ and $\{ \vert 3\ra_2,..., \vert d
\ra_2\}$ respectively. The nondegenerate part of these operators is chosen
such that they yield maximal violation of the Clauser, Horne, Shimony and
Holt (CHSH) inequality [7] for the singlet state $\vert S_{12}\ra$, that is
$$
\la S_{12}\vert AB+AB'+A'B-A'B'\vert S_{12}\ra=2\sqrt 2.\eqno(5)
$$

Let us see now what happens if we start with an ensemble of pairs of
particles in a Werner state $W^{(d\times d)}$ and subject each particle to
the measurements described above. According to the results obtained in the
measurements of P and Q the original ensemble splits into four subensembles
(corresponding to the results $\{0,0\}$, $\{0,1\}$, $\{1,0\}$ and
$\{1,1\}$). The most important point is that if the initial ensemble is
classical, behaving according to a hidden variables model, than each of
these four subensembles is classical. But then we get a contradiction with
the quantum mechanical predictions. Indeed, the ensemble corresponding to
P=1 and Q=1 is described by
$$
W'={1\over N}PQWQP=
{{2d}\over{2d+4}}\bigl({1\over{2d}}I^{(2\times 2)}+\vert S_{12}\ra
\la S_{12}\vert\bigr),\eqno(6)
$$
where N is a normalization factor and and $I^{(2\times 2)}$ is a $2\times
2$ identity matrix acting in the $\{\vert 1\ra_1,\vert
2\ra_1\}\otimes\{\vert 1\ra_2,\vert 2\ra_2\}$subspace and zero in rest. In
this state the CHSH inequality is violated. Indeed,
$$
Tr W'(AB+AB'+A'B-A'B')={{2d}\over{2d+4}}2\sqrt2\geq 2
{}~~~~for~~~~d\geq5.\eqno(7)
$$

In conclusion, although a local hidden variables model can simulate all the
correlations which arise when only a single, ideal measurement is performed
on each of the two particles, such a model cannot account for the
correlations which arise when two consecutive measurements are performed on
each particle.
\bigskip

To better understand the above result, let us compare the case in which the
measurements of A or A' and B or B' take place directly on the original
state W and the case when P and Q were measured first. According to quantum
mechanics the correlations between the outcomes of A, A', B and B' are the
same in both cases, as all these operators commute with P and Q.  Still, a
local hidden variables model can simulate these correlations in the first
case but not in the second one.  This happens because for a hidden
variables model the two situations are dramatically different. Suppose
first that one of the operators A or A' is measured on particle 1 directly
in the initial state. Then the particle has from the beginning all the
information about the detailed question which is asked about the $\{\vert
1\ra_1,\vert 2\ra_1\}$ subspace and it can use this information to avoid
``unpleasant" questions. Indeed, quantum mechanics imposes that a
measurement of A yields the outcome 0 with the same probability as a
measurement of A' (as the corresponding eigenspaces are identical). In a
local hidden variables model this means that
$$
\int d\lambda\rho(\lambda)P_1(A=0,\lambda)=
\int d\lambda\rho(\lambda)P_1(A'=0,\lambda),\eqno(8)
$$
where $\lambda$ is the hidden variable, $\rho(\lambda)$ is the probability
distribution of $\lambda$ over the initial ensemble and $P_1(A=0,\lambda)$
and $P_1(A'=0,\lambda)$ are the probabilities that particle 1 gives the
answers A=0 and A'=0 respectively. But from eq. (8) it {\it does not}
follow that
$$
P_1(A=0,\lambda)=P_1(A'=0,\lambda).\eqno(9)
$$
The same is true for the sum of probabilities to obtain
the results 1 and -1, that is,
$$ \int
d\lambda\rho(\lambda)\bigl[P_1(A=1,\lambda)+P_1(A=-1,\lambda)\bigr]= \int
d\lambda\rho(\lambda)\bigl[P_1(A'=1,\lambda)+P_1(A'=-1,\lambda)\bigr],\eqno(10)
$$
but in general
$$ P_1(A=1,\lambda)+P_1(A=-1,\lambda)\ne
P_1(A'=1,\lambda)+P_1(A'=-1,\lambda).\eqno(11) $$
The meaning of these relations is that particle 1, in some pairs of the
original ensemble, might have a local hidden variable according to which if
it is subjected to a measurement of A, say, it will yield one of the
answers 1 or -1, but if it is subjected instead to a measurement of A', it
will yield 0. Analogously, similar behaviour could characterize also
particle 2 in some of the pairs. This freedom to choose between $\pm1$ and
$0,$ depending on the measurement to which the particles are subjected, is
vital for the success of Werner's local hidden variables model, as he
himself remarked. (Actually in Werner's model only particle 1 in some of
the pairs uses this freedom, while particle 2 is less sophisticated.)  On
the other hand, if we first subject particle 1 to a measurement of P, the
particle commits itself to yield, in a subsequent measurement of either A
or A', one of the outcomes $\pm1$ or $0$ {\it before} knowing which of
these measurements will actually be performed. Indeed, $P=1$ forces the
particle to yield 1 or -1 subsequently no matter if A or A' is measured;
the outcome $0$ is no longer a valid option. Similarly, $P=0$ forces the
outcome $0$ in a subsequent measurement of either A or A'. At this point no
local hidden variables model could simulate the quantum mechanical
behavior.
\bigskip
Another way of understanding the role of the measurements of P and Q is the
following: they are used to select a subensemble of pairs from the original
ensemble in a way {\it independent} of the measurements which are used to
test it. Because of this independence one could apply the CHSH inequality
directly to this subensemble and conclude that the larger than 2 value of
the CHSH correlation predicted by quantum mechanics for this subensemble
(eq. 7) is incompatible with a local hidden variables model. On the other
hand, one could think of measuring A or A' and B or B' directly on the
original ensemble and then selecting the subensemble of pairs for which
both particle 1 and particle 2 have yielded $\pm1$. But in a local hidden
variables model such as Werner's, the subensemble selected this way {\it
depends} on the measurements used to test it. Indeed, suppose that particle
1 in a particular pair has a local hidden variable according to which, if
it is subjected to a measurement of A, it will yield, say, 1, while to a
measurement of A' it will yield 0. Then if it happens that particle 1 is
subjected to a measurement of A, the pair might be included in the selected
subensemble (depending on a $\pm1$ outcome yielded by particle 2), while if
it happens that particle 1 is subjected to a measurement of A' the pair
will not be selected. For such test-dependent subensembles Bell's
inequalities cannot be applied directly and, as Werner shows, values larger
than 2 for the CHSH correlations can be consistent with local hidden
variables models. Other, much more famous, examples of test-dependent
subensembles and apparent Bell's inequality violations are the ``detector
efficiency" problem, and the ``space distribution" problem raised by
E. Santos [8]. Another example of this type, probably the simplest
possible, has been given by D. Mermin [9].
\bigskip
There are, of course, many open questions left. In the example presented in
this letter the nonlocality of this special state was revealed by
considering a sequence of ideal measurements. But it might be that a single
but {\it non-ideal} measurement performed on each particle suffices. For
example, we could bring along each of the original two particles some
supplementary quantum system (called sometimes ``ancilla") and then perform
on each side an ideal measurement on the original particle and the
corresponding ancilla together. Such a measurement can, in principle yield
more outcomes that d, the dimension of the hilbert space of the original
particle. This scheme is known in the literature as the implementation of a
positive valued operator measure (POVM) and it is known that in certain
situations it can yield more information about a quantum system than an
ideal measurement performed only on the system itself. Using non-ideal
measurements is obviously necessary in the case of Werner's $2\times 2$
matrix.  Indeed, in this case the strategy of subjecting each system to a
sequence of ideal measurements does not work, as after the first
measurement the systems are left in a direct product state and therefore
any subsequent measurements yield classical correlations.

Finally, an open question is what is the class of classical states. Is it
the case that only the mixtures which are decomposable as a sum of direct
product states are classical and all the others are not, although
superficially they seem to be? As a first test case one should probably
attack the remaining Werner type matrices ($d\leq4$).

It is a pleasure for me to thank Lev Vaidman and the anonymous referee B of
my paper [6] for raising questions which led to this work, A. Zeilinger for
some very illuminating remarks and also L. Hardy, A. Mann, A. Peres,
M. Revzen, D. Rohrlich and B. Tsirelson (Cirel'son) for the very
interesting discussions we had.

References

1) J. S. Bell, Physics 1, 195, (1964).

2) N. Gisin, Phys. Lett. A 145, 201,  (1991); S. Popescu and D. Rohrlich,
Phys. Lett. A 166, 293, (1992).

3) S. Braunstein, A. Mann  and M. Revzen, Phys. Rev. Lett. 68, 3259,
(1992); A. Mann, K. Nakamura and M. Revzen, J. of Phys. A L851, (1992).

4) For an extensive reference list see L. Khalfin and B. Tsirelson,
Found. of Phys. 22, 879, (1992).

5) R. F. Werner, Phys. Rev. A 40, 4277, (1989).

6) S. Popescu, Phys. Rev. Lett. 72, 797, (1994).

7) J. Clauser, M. Horne, A. Shimony and R. Holt, Phys. Rev. Lett. 23,
880, (1969).

8) E. Santos, Phys. Rev. Lett. 66, 1388, (1991); Phys. Rev. Lett. 68,
2702, (1992); Phys. Rev. A 46, 3646, (1992).

9) D. Mermin, ``Boojums All the Way Through", Cambridge University
Press, NY (1990) pps. 167-169.

\end